\documentclass[sigconf]{acmart}
\AtBeginDocument{%
  }

\setcopyright{acmlicensed}
\copyrightyear{2026}
\acmYear{2026}
\acmDOI{XXXXXXX.XXXXXXX}
\acmConference[Conference]{}{2026}{USA}
\acmISBN{978-1-4503-XXXX-X/2018/06}

\usepackage{multirow}
\usepackage{subfigure}

\begin{document}
\title{Decoding the Hook: A Multimodal LLM Framework for Analyzing the Hooking Period of Video Ads}

\author{Kunpeng Zhang}
\email{kpzhang@umd.edu}
\affiliation{%
  \institution{University of Marland, College Park}
  \city{College Park}
  \state{Maryland}
  \country{USA}
}

\author{Poppy Zhang}
\email{poppyzhang@meta.com}
\affiliation{%
  \institution{Meta Platforms, Inc.}
  \city{New York}
  \country{USA}}

\author{Shawndra Hill}
\email{shawndrahill@meta.com}
\affiliation{%
  \institution{Meta Platforms, Inc.}
  \city{New York}
  \country{USA}
}

\author{Amel Awadelkarim}
\email{ameloa@meta.com}
\affiliation{%
  \institution{Meta Platforms, Inc.}
  \state{California}
  \country{USA}
}

\renewcommand{\shortauthors}{Zhang et al.}

\begin{abstract}
Video-based advertisements have become an important medium for brands to engage consumers, with social media platforms leveraging extensive user data to optimize ad delivery and enhance engagement. An under-explored aspect of video ad effectiveness is the initial ``hooking period" — the first three seconds that capture viewer attention and influence subsequent engagement metrics. Analyzing the factors that drive performance during this brief time frame is challenging due to the multimodal nature of video content, which integrates visual, auditory, and textual elements. Traditional analysis methods often fall short in capturing the nuanced interplay of these components, necessitating advanced frameworks for comprehensive evaluation.

This study introduces a framework that employs transformer-based multimodal large language models (MLLMs) to dissect the hooking period of video advertisements. It tests two different frame sampling strategies — uniform random sampling and key frame selection — to ensure a balanced and representative acoustic feature extraction, capturing the full spectrum of design elements. The hooking video is processed by state-of-the-art MLLMs to generate descriptive analyses of the ad's initial impact, which are then distilled into coherent topics using BERTopic for high-level abstraction. Additionally, the framework integrates additional features such as audio attributes and aggregated ad targeting information, enriching the feature set for subsequent analysis.

Empirical validation on large-scale real-world data from social media platforms demonstrates the efficacy of our framework, revealing correlations between hooking period features and the key performance metrics like conversion per investment. Our results highlight the practical applicability and predictive power of the proposed approach, offering valuable insights for optimizing video ad strategies. This study advances the state-of-the-art in video ad analysis by providing a comprehensive and scalable methodology for understanding and enhancing the initial moments of video advertisements.
\end{abstract}


\keywords{Video ads, Multimodal LLM, Hooking period, Conversion, Feature extraction}

\settopmatter{printacmref=false}
\maketitle

\section{Introduction}
In the rapidly evolving digital landscape, video-based advertisements have emerged as a pivotal medium for brands to engage consumers. Social media platforms have capitalized on this trend, leveraging vast amounts of user data to optimize ad delivery and enhance user engagement \cite{zhangweb2024}. Understanding the elements that contribute to the effectiveness of video ads is paramount for both advertisers and platform providers. Specifically, the initial moments of an advertisement—the so-called "hooking period" comprising the first three seconds—are important in capturing viewer attention and influencing subsequent engagement metrics. Despite the significance of this brief yet impactful timeframe, comprehensively analyzing the factors that drive ad performance during the hooking period remains a formidable challenge.

Video advertisements are a cornerstone of digital marketing strategies, offering dynamic and immersive experiences that static ads cannot match \cite{krishnanimc2013}. The ability to convey compelling narratives, evoke emotional responses, and showcase products or services within a limited timeframe makes video ads exceptionally effective \cite{kangjcb2020}. For social media platforms, which facilitate a considerable amount of ad impressions daily, optimizing ad performance directly correlates with user satisfaction and platform growth \cite{sagtaniwsdm2024}. The hooking period is particularly influential, as it determines whether a viewer continues watching the ad or scrolls past it. A well-crafted hooking period can significantly enhance metrics such as impressions, conversion per investment (CPI, investment referring to advertisement budget), and user engagement rates, thereby delivering substantial returns on advertising investments \cite{vedulaicdm2017}.

However, the complexity of video content, which encompasses visual elements, audio cues, and temporal dynamics, poses significant challenges for analysis. Traditional methods often rely on manual annotation or simplistic feature extraction techniques that fail to capture the nuanced interplay of multimodal elements within the hooking period \cite{cai2025lovrbenchmarklongvideo}. Consequently, there is a pressing need for advanced analytical frameworks that can automatically extract and interpret the important features of video ads, thereby informing the design and optimization of more effective advertising strategies.

The task of dissecting and understanding the hooking period of video ads involves several intricate challenges. First, the multimodal nature of video content—integrating visual, auditory, and textual information—necessitates sophisticated models capable of processing and interpreting diverse data types simultaneously. Traditional machine learning approaches may struggle with this complexity, often requiring separate processing pipelines for different modalities, which can lead to fragmented and less coherent feature representations \cite{baltrusaitisieee2019}. Second, existing methodologies in video ad analysis predominantly focus on either broad content classification or superficial feature extraction \cite{guomm2021}, inadequately addressing the specific dynamics of the initial three seconds. These methods often overlook the subtle yet impactful design elements that characterize successful hooking periods, such as emotional appeal, visual aesthetics, interactivity, and challenges posed to the viewer \cite{cokerijrm2021}. Moreover, the temporal aspect of video content, where the sequence and timing of elements play an important role in viewer retention, is frequently under-explored \cite{ye2025rethinking}. Another significant challenge lies in linking the extracted features of the hooking period to concrete performance metrics. While various studies have attempted to correlate ad attributes with outcomes like impressions and engagement rates, establishing a robust and predictive relationship remains elusive. This gap is exacerbated by the sheer volume of ad data, which demands scalable and efficient analytical frameworks capable of handling large-scale datasets without compromising on the granularity of insights.

Addressing these challenges necessitates a multifaceted approach that leverages the latest advancements in machine learning, particularly in the realm of multimodal large language models (MLLMs) \cite{jin2024mllm}. Our proposed framework harnesses the power of transformer-based architectures \cite{vaswani2023attentionneed} to comprehensively analyze the hooking period of video ads, extracting and interpreting key features that drive performance metrics. The process begins with the conversion of each video ad into a sequence of frames, employing two distinct sampling strategies: uniform random sampling and key frame selection \cite{keyframe2025}. Uniform random sampling ensures a representative distribution of frames across the entire hooking period, while key frame selection targets frames that encapsulate significant visual or narrative shifts. This dual approach facilitates a more nuanced understanding of the temporal dynamics within the hooking period. Subsequently, the extracted frames are fed into state-of-the-art MLLM, which are adept at processing and interpreting multimodal data. These models generate text-based reasoning descriptions that encapsulate the design methodologies employed in the hooking period. For instance, the model may identify elements related to emotional appeal, visual aesthetics, interactivity, or viewer challenges. Such descriptive analyses provide a structured representation of the ad's initial impact, facilitating deeper insights into the factors that influence viewer engagement. To further distill the rich textual descriptions generated by the MLLMs, we employ BERTopic \cite{grootendorst2022bertopic}, a topic modeling technique that leverages transformer-based embeddings to identify coherent topics within large text corpora. By summarizing the methodology descriptions into a few salient topics, we obtain a high-level overview of the prevalent design strategies within the hooking period. This topic-level abstraction enables the integration of qualitative insights with quantitative performance metrics. In addition to the multimodal features extracted from the hooking period, our framework incorporates auxiliary features such as audio attributes and aggregated-level ad information (e.g., targeting, ad placement). By synthesizing these diverse data sources, we construct a comprehensive feature set that encapsulates both the intrinsic qualities of the ad content and the contextual factors influencing its performance. Finally, we employ predictive modeling techniques to establish the relationships between the extracted features and key performance metrics, including impressions, conversion per investment, and engagement rates. By training models on this enriched feature set, we aim to achieve improved predictive performance, thereby enabling more accurate forecasting of ad success based on the characteristics of the hooking period.

In summary, this paper presents a comprehensive and innovative approach to understanding the hooking period of video ads through the lens of multimodal large language models. By addressing the existing research gaps and introducing novel analytical techniques, we contribute valuable insights and methodologies that advance the state-of-the-art in video ad analysis and optimization. Specifically, we contribute to the literature as follows.
\begin{enumerate}
    \item Innovative Multimodal Analysis Framework: We introduce a framework that leverages multimodal large language models to extract and interpret key features from the hooking period of video ads. This approach effectively integrates visual, auditory, and textual data, providing a comprehensive understanding of the elements that drive ad performance.
    \item Frame Sampling Strategies for Acoustic Feature Extraction: By testing both uniform random sampling and key frame selection, our method ensures a balanced and representative extraction of frames, capturing the full spectrum of design elements within the hooking period. This enhances the robustness and granularity of the feature extraction process.
    \item Integration of Auxiliary Features: Our framework seamlessly incorporates additional features such as audio attributes and aggregated ad information, enriching the feature set and enhancing the predictive capabilities of the model. This holistic approach ensures that both content-specific and contextual factors are accounted for in the analysis.
    \item Empirical Validation on Real-World Data: Utilizing five categories of real-world data from a social platform, we demonstrate the efficacy of our interpretable framework through extensive experiments. The results reveal insightful findings regarding the relationship between hooking period features and ad performance metrics, showcasing the practical applicability and predictive power of our approach.
\end{enumerate}

\section{Related Work}
\label{sec:relatedwork}
We review the relevant research from the following two perspectives, including the use of multi-modal large language models for video understanding and ads performance prediction.

\textbf{Video understanding} 
Recent advances in video understanding have significantly benefited from deep learning techniques, particularly leveraging architectures such as convolutional neural networks (CNNs) and Transformers. Early methods include extending 2D CNNs into 3D variants, such as C3D \cite{tran2015} and I3D \cite{carreira2017}, to effectively capture temporal information within videos, thereby achieving significant improvements in tasks like video classification and event detection. More recently, Transformer-based models, including Video Vision Transformer (ViViT) \cite{arnab2021} and TimeSformer \cite{bertasius2021}, have further advanced the state-of-the-art by explicitly modeling spatio-temporal relationships using self-attention mechanisms. These approaches excel at modeling long-range dependencies across video frames, enhancing their capability to capture complex semantic contexts.

However, despite these advancements, accurately interpreting content effectiveness in video advertisements remains challenging, largely due to multimodal dynamics involving visual, audio, and textual cues \cite{gao2020,lei2021}. Recent studies have explored multimodal fusion methods that integrate different modalities to better predict ad performance metrics. For example, multimodal Transformer-based models jointly encoding visual, audio, and textual features have demonstrated promising results in unerstanding viewer reactions and improving content personalization \cite{gabeur2020, lee2021}. To sum up, optimizing such models for video ad performance prediction remains an active research area, particularly in identifying effective methods for integrating multimodal data and interpreting the complex interactions among multiple modalities.

\textbf{Ads performance prediction}
Ad performance prediction using images or videos has gained considerable attention in recent years due to the rich content embedded in these mediums. Traditional methods of predicting ad performance mainly relied on text-based features and user interactions, but the inclusion of images or videos allows for a deeper analysis of the content itself. Researchers have  leveraged advanced deep learning techniques, particularly Convolutional Neural Networks (CNNs) and Recurrent Neural Networks (RNNs), to successfully extract features from visual, audio, and textual data that are correlated with key performance metrics such as click-through rates (CTR), conversion rates, and overall user engagement \cite{chen2016mm}. They include color, objects, and facial expressions in images, motion and scene changes in videos, as well as text-based features. Moreover, recent advancements in multimodal learning have enabled the integration of visual, audio, and textual data for even more accurate performance forecasting \cite{Gharibshah2021}. Such a cross-modal approach captures the complex relationships between visual appeal and user responses, considering factors like emotional tone, product presentation, and visual aesthetics \cite{jun2020}. As digital advertising continues to evlove, these image- and video-based prediction models are expected to play a key role in optimizing ad strategies, ensuring that marketers can more effectively target their audiences and maximize return on investment (ROI) \cite{lee2021access}.

\vspace{-7mm}
\section{Methodology}
\label{sec:method}
\begin{figure*}[ht]
  \centering
  \includegraphics[width=0.9\linewidth]{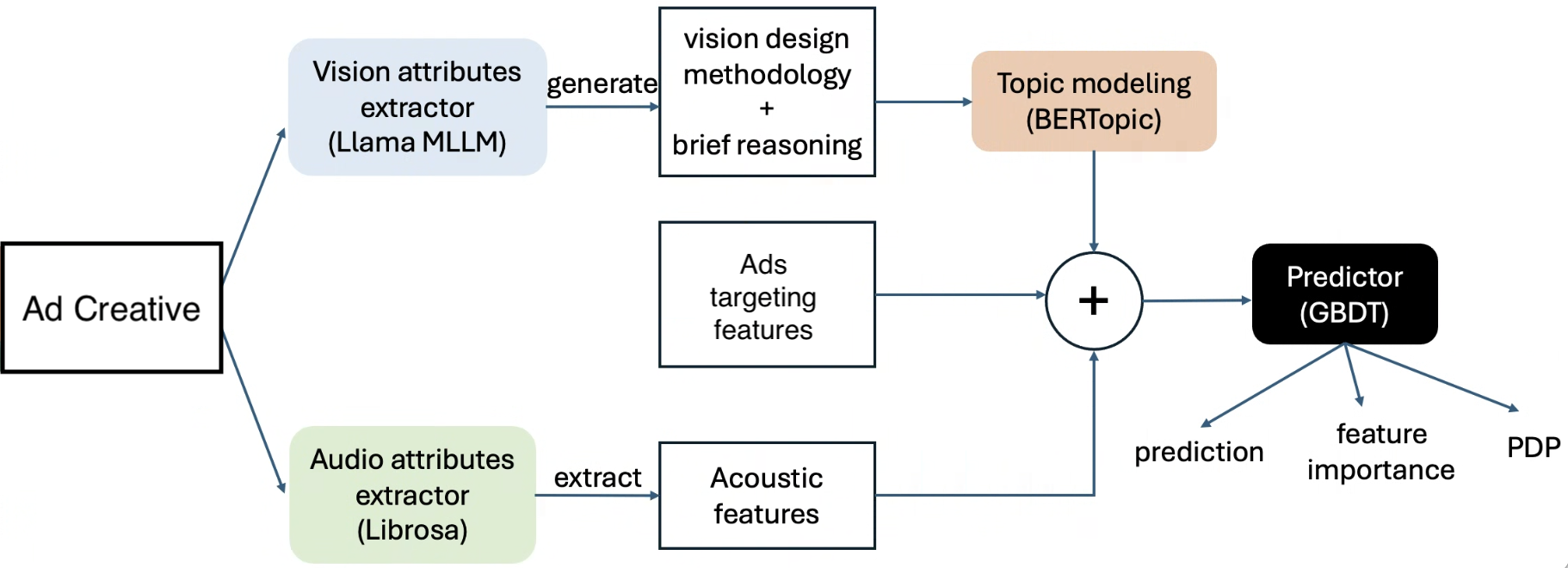}
  \caption{Overview of our proposed framework: multimodal LLM-based video ad understanding (MLLM-VAU)}\label{fig:model}
\end{figure*}

In this section, we present a detailed description of our proposed method, MLLM-VAU (Multimodal LLM-based Video Ad Understanding). The overall framework of MLLM-VAU is illustrated in \figurename~\ref{fig:model}. We first introduce an overview of the preliminaries in Section~\ref{sec:preliminary}. Next, we detail the four core components of MML-VAU: the video processor (Sec.~\ref{sec:processor}), prompt-based vision insights extractor (Sec.~\ref{sec:extractor}), audio attributes extractor (Sec.~\ref{sec:audio}), and the predictive analyzer (Sec.~\ref{sec:predictor}).

\subsection{Preliminary}
\label{sec:preliminary}
Let $\mathcal{V} = v^1, v^2, \cdots, v^n$ denote a set of videos in a specific industry vertical (e.g., Ecommerce). Each video $v^i = \{\mathcal{F}^i, \mathcal{A}^i, \mathcal{T}^i\}$ has three components, where $\mathcal{F}^i$ represents a sequence of $m$ image frames extracted from the hooking period of the video, $\mathcal{F}^i=\{\mathcal{I}^i_1, \mathcal{I}^i_2, \cdots, \mathcal{I}^i_m\}$, $\mathcal{A}^i$ is the audio component in the hooking period, and potentially $\mathcal{T}^i$ represents the text if the hooking period has human speech rather just background music. $\mathcal{T}^i$ can also be some textual descriptions of the video such as the title or the short summary. The objective of MLLM-VAU is to understand which design feature from three components have an impact on some predefined metrics such as performance (e.g., click-through rate, impression, conversion per investment) or engagement (e.g., likes, shares). In recent practice, researchers and practitioners mainly design various methods to extract features from a video. Each individual feature is often a well-trained model based on some manually labeled datasets. In addition, there exist many approaches that convert a video into an embedding upon which a predictive model is then built. One of the major drawbacks is lack of interpretability due to its nature of `black-box', which does not provide any guidance for advertisers and thus has limited practical value.

\subsection{Video Processor}
\label{sec:processor}
To harness the capabilities of multimodal large language models (MLLMs) for analyzing the hooking period of video advertisements, a sophisticated video processing framework is essential. This framework is designed to preprocess raw video data by extracting sequences of image frames, audio attributes, and generating corresponding textual descriptions. These multimodal inputs provide a comprehensive foundation for subsequent analysis, ensuring that the intricate interplay between visual, auditory, and textual elements within the hooking period is effectively captured and interpreted.

The initial phase of the video processing framework involves the extraction of image frames and audio attributes from each video advertisement. Image frames are sampled to represent the temporal dynamics of the first three seconds—the hooking period—of the advertisement. Concurrently, the audio stream is processed to extract relevant attributes such as volume levels, frequency components, and temporal variations, which are important for understanding the auditory appeal of the ads. Additionally, automatic speech recognition (ASR) systems are employed to transcribe any spoken content if it exists, thereby generating textual descriptions that complement the visual and auditory data. This multimodal preprocessing ensures that the MLLMs receive a rich and diverse set of features, enabling a more nuanced analysis of the ad's initial impact.

\textbf{Two Frame Sampling Strategies} A pivotal component of the video processing framework is the selection of frames that best represent the hooking period. We have developed and implemented two distinct frame sampling strategies: \textit{random sampling} and \textit{key frame selection}. Random sampling involves selecting frames at a uniform interval throughout the hooking period without any bias towards specific content changes or significant visual events. It can be formalized as uniformly and independently selecting a subset of $m$ frames from the total of $K$ frames, where $K$ is determined by the frames per second (fps). For example, the hooking period has 90 frames if fps=30. $\mathcal{F}^i=\{\}$. This ensures a broad and unbiased representation of the entire hooking period, capturing a diverse array of visual and auditory elements. Additionally, it is straightforward to implement and computationally efficient, making it suitable for processing large volumes of video data rapidly. However, it may miss key moments or significant transitions within the hooking period that are important for capturing the ad's impact. This approach does not account for the semantic or emotional significance of specific frames, potentially diluting the relevance of the extracted features. 

On the other hand, key frame selection focuses on identifying and extracting frames that encapsulate significant visual or narrative shifts within the hooking period. The process involves three key steps: 
\begin{itemize}
    \item Calculating the difference $D_i$ between consecutive frames ($D_i = 1 - SSIM(I_i, I_{i+1})$, $I_i$ is the $i^{th} frame$) using a suitable metric, such as the Structural Similarity Index Measure (SSIM).
    \item Determine a threshold $\tau$ to identify significant changes: $\tau=\alpha*max(D_i, D_2, \cdots, D_{K})$, where $\alpha$ is a scaling factor (e.g., 0.5).
    \item Select frames where the difference exceeds the threshold: $\mathcal{F}=\{I_i | D_i > \tau\}$. To ensure temporal diversity, impose a minimum interval $\Delta t$ between consecutive key frames: $\mathcal{F}=\{I_i | D_i > \tau$ and $i-j 
    \geq\Delta t~ \forall$ selected frame, $j<i\}$.
\end{itemize}
 This strategy leverages algorithms designed to detect changes in scene composition, motion intensity, and other salient features that indicate pivotal moments in the advertisement. As we can see that this strategy requires additional processing overhead, as it involves running algorithms to detect significant changes or events within the video. It may also introduce bias by prioritizing certain types of changes or events, potentially overlooking other relevant aspects of the hooking period. To sum up, the frame selection between random sampling and key frame selection presents a trade-off between computational efficiency and the depth of contextual representation.

\subsection{Prompt-based Vision Insights Extractor}
\label{sec:extractor}
To extract nuanced insights from each video advertisement, we employ prompt-based multimodal large language models (MLLMs). Specifically, we designed an extractor: vision design methodology extractor.

\textbf{Vision Design Methodology Extractor} This employs Llama Multimodal Model \footnote{\url{https://www.llama.com/}}, which leverages the advanced capabilities of transformer-based architectures to interpret and analyze the complex interplay among a sequence of selected image frames within the hooking period of video ads. By designing tailored prompts, we guide the model to assess the primary engagement strategies employed in each advertisement, thereby facilitating a deeper understanding of the factors that drive ad performance.

The effectiveness of MLLMs in extracting meaningful insights heavily relies on the design of the input prompts. Our prompts are meticulously crafted to include both the title and a detailed textual description of the video ad, providing comprehensive context for the model. The prompt is specifically tailored to elicit the model's assessment of the primary engagement strategy used as the hook in the first three seconds of the ad. The prompt we used is below.
\begin{verbatim}
        After examining the video and text 
        advertisement titled "{ad title text}" 
        with the body texts "{ad body text}", 
        determine the primary method used by the 
        advertiser to engage the audience. Base 
        your selection on the actual content of 
        the advertisement without making 
        assumptions or interpretations. 

        Respond using the JSON format.
        **JSON Response Format:**
        {{
        "methodology": "Methodology chosen by 
        the advertiser", "rationale": "Provide 
        a concise explanation based on specific 
        elements observed in the advertisement 
        that supports why this option best 
        represents the primary engagement 
        strategy used."
        }}
\end{verbatim}

This prompt structure ensures that the model receives sufficient contextual information to make an informed assessment. By explicitly asking for the identification of engagement strategies and a corresponding rationale, we obtain both categorical and explanatory outputs that enrich the feature set used for predictive modeling. Upon processing the crafted prompt, MLLM generates a structured output that includes both the assessment of the primary engagement strategy and a textual rationale explaining the reasoning behind the assessment. This dual-output mechanism provides a clear and interpretable understanding of the strategies used in the hooking period, allowing for more granular analysis and feature extraction.

Employing MLLM for feature extraction offers several distinct advantages over traditional methods. First, the adaptability of prompt-based interactions allows for enhanced contextual understanding. Because LLMs, by design, are trained on vast corpora of data encompassing diverse contexts and can interpret image frames through a lens that incorporates not only visual features but also contextual and semantic information. Second, LLMs offer unparalleled flexibility through prompt-based interactions, allowing for the extraction of a wide array of features without the need for retraining or extensive reconfiguration. By designing specific prompts, researchers can tailor the feature extraction process to focus on particular aspects of the image that are hypothesized to influence advertisement performance. Third, traditional ML approaches often necessitate significant manual effort in feature engineering, involving the selection, transformation, and combination of various visual attributes to create meaningful features. This process is not only time-consuming but also prone to human bias and may miss subtle yet impactful features. LLM-based feature extraction automates much of this process by leveraging the model's inherent ability to understand and synthesize information based on the prompts provided.

To further distill the rich textual rationales generated by Llama, we employ BERTopic, a topic modeling technique that leverages transformer-based embeddings to identify coherent topics within large text corpora. This step transforms the qualitative rationales into a structured set of latent topics, facilitating their integration into subsequent predictive models. The latent topics derived from BERTopic are integrated with other multimodal features such as audio attributes and aggregated ad information. This comprehensive feature set encapsulates both the qualitative insights from the engagement strategies and the quantitative aspects of the video content. Overall, employing prompt-based MLLMs in conjunction with BERTopic offers several advantages: (i) Contextual Understanding: The ability to process and interpret multimodal inputs allows for a more nuanced analysis of video content compared to traditional feature extraction methods. (ii) Detailed Insights: The generation of both categorical assessments and rationales, followed by topic modeling, provides a layered understanding of the underlying engagement strategies. (iii) Scalability: Automated prompt-based analysis and topic modeling enable the processing of large-scale video datasets efficiently. (iv) Enhanced Interpretability: Latent topics summarize complex rationales into coherent themes, facilitating easier interpretation and integration into predictive models.

\subsection{Audio Attributes Extractor}
\label{sec:audio}
Beyond visual stimuli, acoustic elements also play a pivotal role in shaping the viewer's perception and emotional response. This study emphasizes the extraction and analysis of specific acoustic features during this hooking period to enhance the predictive accuracy of advertisement performance models. The acoustic features we focus on include decibels (dB), jitter, tempo, degree of dynamic pitch (DDP), pitch (maximum, minimum, mean), power, peak, and shimmer. These features collectively provide a comprehensive understanding of the audio dynamics that contribute to an advertisement's effectiveness. Details about each acoustic feature is below.
\begin{itemize}
    \item Decibels (dB): Decibels measure the loudness of the audio signal. In the context of advertisements, variations in volume can influence attention and emotional intensity. A sudden increase in dB may be used to highlight key moments or transitions, thereby enhancing the memorability of the ad.
    \item Jitter: Jitter refers to the frequency variation or instability in the audio signal. High jitter levels can indicate a more dynamic or erratic audio pattern, which may be used to convey excitement or urgency. Conversely, low jitter can create a sense of calmness and stability, aligning with the intended message of the advertisement.
    \item Tempo: Tempo denotes the speed or pace of the audio track. A faster tempo can energize viewers and create a sense of urgency, while a slower tempo may evoke relaxation and contemplation. Understanding the tempo during the hooking period helps in assessing how the audio pace aligns with the visual content to engage viewers effectively.
    \item Degree of Dynamic Pitch (DDP): DDP captures the variability and changes in pitch over time. This feature helps identifying tonal shifts that can signal different emotional states or transitions within the advertisement. By analyzing DDP, we can better understand how pitch variations contribute to maintaining viewer interest during the initial moments.
    \item Pitch (Maximum, Minimum, Mean): Pitch analysis provides insights into the fundamental frequency of the audio. The maximum and minimum pitch values, along with the mean pitch, help in characterizing the overall tonal range and emotional tone of the advertisement. For instance, higher pitches may be associated with excitement or positivity, while lower pitches might convey seriousness or authority.
    \item Power: Power measures the energy of the audio signal, reflecting its overall strength and presence. High power levels can make an advertisement more impactful and memorable, whereas lower power may be used to create subtlety or intimacy. Analyzing power during the hooking period helps in understanding how audio intensity influences viewer engagement.
    \item Peak: Peak detection identifies the highest amplitude points in the audio signal. Peaks often correspond to key moments or emphases in the advertisement, such as a dramatic sound effect or a vocal highlight. Recognizing these peaks is essential for assessing how audio cues are used to draw attention and reinforce the advertisement's message.
    \item Shimmer: Shimmer quantifies the amplitude variation or instability in the audio signal. This feature is indicative of subtle fluctuations in volume that can add expressiveness and nuance to the audio track. Shimmer analysis helps in evaluating how these subtle variations contribute to the overall emotional and psychological impact of the advertisement.
\end{itemize}

\subsection{Predictor}
\label{sec:predictor}
To understand ad performance, we integrate the rich features extracted from the hooking period of video ads (e.g., those discussed above) with ad characteristics. The hooking period features include visual design methodology regarding engaging audiences derived from MLLM , and acoustic characteristics, while the aggregated user data are gender, age bucket, advertiser size, zip code, and others. We then link these features to the ad performance metric, conversion per investment (CPI), using Gradient Boosting Decision Tree (GBDT)\cite{gbdticml,pdp2001}. By training on historical ad performance data, we are able to identify and quantify the correlations between specific features and performance metrics. This not only enhances the predictive accuracy but also provides valuable insights into which aspects of the hooking period -- be it certain visual compositions or acoustic dynamics — are most influential in driving successful ad engagements. Ultimately, this approach enables advertisers to optimize their content strategically, leveraging data-driven insights to enhance the effectiveness and impact of their campaigns.

\section{Experiments and Results}
\label{sec:results}
In this section, we first introduce the data and experimental setup (see Sec. \ref{sec:data}). Then we show the performance comparison results of our method with two baselines: one strongly designed one and one intuitively simple one in Sec \ref{sec:performance}. Major findings are detailed in Sec. \ref{sec:findings}, including the topics summarizing the hook responses into different categories, overall important features that affect the CPI, and the partial dependency plots. Finally, we showcase several examples to reinforce the practical values of our study in the Appendix.

\subsection{Data and Experimental Settings}
\label{sec:data}
In this study, we focus on advertisers whose ads with one video have a minimum spend of certain amount for the first day and for the 56 days since their launch. The descriptive statistics of our final dataset are described in Table \ref{tab:stats}. All videos are in the format of MP4. The median video length is 29 seconds.
\begin{table}[ht]
    \centering
    \caption{Descriptive statistics of datasets. Note that ``CPG" refers to ``Consumer Packaged Goods".}
    \label{tab:stats}
    \begin{tabular}{l|c|c|c}
        \toprule
         & \# of videos  & [min,max,mean] & \% of videos \\
        Category & (Binned) & CPI & w/ audios \\
        
        \midrule
        Ecommerce    & 100k - 150k    & [0, 52.60, 0.028]     & 94.72     \\ 
        Healthcare    & 25k-50k     & [0, 33.20, 0.045]     & 87.11    \\ 
        CPG    & 75k-100k     & [0, 37.25, 0.092]     & 91.87      \\ 
        Automobile    & 10k-25k     & [0, 56.25, 0.219]     & 87.43    \\ 
        Entertainment    & 10k-25k    & [0, 13.31, 0.048]     & 95.43     \\ 
        \bottomrule
    \end{tabular}
\end{table}

\textbf{Implementation Details\footnote{Sample data and all code can be provided upon request.}} We use the Llama MLLM as the vision design methodology extractor. For the GBDT model, we use the grid search to tune the hyperparameters. The final optimal setting is: \#of trees=740, tree depth=12, learning rate=0.0764, minimum number of data points in a node that is required for the node to be split further=50, sample size = 0.86 (the proportion of data that is exposed to the fitting routine), and the others are default values. The acoustic features are extracted using the python package librosa.\footnote{https://librosa.org/doc/latest/index.html} Note that all experiments can be done within a few hours (e.g., about 20 hours for the Ecommerce) in total, including feature extraction, training, and testing. 

\textbf{Benchmark and Metrics} To validate our method, we evaluate its performance and compare it with a strong, carefully designed baseline, widely adopted in video analytics - ViViT \footnote{https://huggingface.co/docs/transformers/en/model\_doc/vivit} and X-CLIP \footnote{https://huggingface.co/docs/transformers/en/model\_doc/xclip}. ViViT is a transformer-based neural network model designed for video analysis that extends Vision Transformers (ViT \cite{vit2021}) by incorporating spatiotemporal attention to effectively capture both spatial and temporal information in video sequences. X-CLIP is a straightforward yet powerful architecture designed to adapt pretrained language-image models for direct video recognition. It comprehends video clips by employing a cross-frame attention mechanism, which explicitly facilitates information exchange across multiple frames. For the efficiency purpose, we uniformly sample 8 frames from the hooking period. We use ViVit/X-CLIP to obtain the embedding of a video, which is fed into GBDT model for CPI prediction along with the same set of variables related to ads targeting. We also benchmark against a simple baseline, which we refer to as a ``junk" predictor. It converts each video hook into a vector by aggregating the pixel values of each frame, upon which the CPI is predicted along with other targeting variables. We use standard evaluation metrics: \textit{R-squared value} ($R^2$) and \textit{mean square error (MSE)} to demonstrate the model fit and the magnitude of the difference between truth and prediction, respectively. 
\begin{table*}[ht!]
    \centering
    \caption{Performance comparison results. Note: The best performance are highlighted in bold.}
    \label{tab:compare}
    \begin{tabular}{l|cccc|cccc}
        \toprule
        \multirow[b]{2}{*}{\textbf{Vertical}} & \multicolumn{4}{c|}{\textbf{R-squared ($R^2$)}} &  \multicolumn{4}{c}{\textbf{MSE}} \\ 
        \cline{2-9}
         & Ours & ViViT \cite{vivit} & X-CLIP \cite{xclip} & ``Junk" & Ours & ViViT & X-CLIP & ``Junk" \\ 
        \hline
        Ecommerce & \textbf{0.21} &  0.08 & 0.20 & 0.08 & \textbf{0.18}  & 0.23 & 1.13 & 0.23  \\ 
        Healthcare & 0.33 &  0.28 & 0.27 & \textbf{0.42} & 0.17  & 0.22 & 0.18  &  \textbf{0.15}\\ 
        CPG & \textbf{0.50}  &   0.34 & 0.07 &  0.27 & \textbf{0.21}  & 0.21 & 0.39 &  0.21\\ 
        Automobile& \textbf{0.66} &  0.57 & 0.46 &  0.11 & \textbf{1.95}  & 2.30 & 2.09  &  3.51\\ 
        Entertainment & 0.25 &  \textbf{0.69} & -1.18 &  0.07  & 0.12  &  \textbf{0.02} & 0.36 &  0.15\\ 
        \bottomrule
    \end{tabular}
\end{table*}

\subsection{Performance Comparison}
\label{sec:performance}
The comparison results are shown in Table \ref{tab:compare}. From the table, we have the following observations. First, our method outperforms both the strong and the weak baselines for Ecommerce, CPG, and Automobile on both \textit{$R^2$} and \textit{MSE}. Second, ViVit achieves the best performance for videos in the vertical of Entertainment. One possible reason is that each entertainment video has a lot more than the fixed number of key frames that are sampled and processed by the Llama MLLM model. ViVit uses all frames in the hooking period, which prevents from losing much information. However, it is worth noting that ViViT is the black-box model and cannot provide any actionable insights to advertisers. Third, the ``Junk'' benchmark does perform the best for the Healthcare category in terms of \textit{$R^2$}, but comparable to our method regarding \textit{MSE}. This could be due to the fact that salient features of videos in this category are primarily demo/products (see Table \ref{tab:top_features}). This indicates that raw pixels of image frames in the hooking period for this vertical might be enough. Finally, X-CLP does not perform well, due to two possible reasons. (i) Many CLIP-based approaches, including X-CLIP, often process videos by extracting frame-level features and then averaging them. This can miss important temporal dynamics, such as motion or action sequences, which are crucial for understanding video content beyond static scenes. (ii) CLIP and similar models are pretrained on large-scale image-text pairs, not video data. This means they may lack the ability to model temporal relationships and actions that are unique to videos.

\subsection{Further Findings}
\label{sec:findings}
In this section, we present top 3 visual design methodologies (topics), top 3 acoustic features correlated to CPI, and their partial dependencies to CPI for all five verticals.
\begin{table}[ht]
\centering
\caption{Top visual and acoustic features that are correlated with CPI by vertical}
\label{tab:top_features}
\begin{tabular}{l|l|l}
    \toprule
    \textbf{Vertical} & \textbf{Visual} & \textbf{Acoustic}\\
    \midrule
    \multirow{3}{*}{Ecommerce} & Topic 6: Interactive content & dB\\
    & Topic 7: Interaction & power \\
    & Topic 10: Connection & maximum pitch\\
    \midrule
    \multirow{3}{*}{Healthcare} & Topic 10: Demo / product & power\\
    & Topic 9: Connection & peak\\
    & Topic 4: Endorsement / celebrity & shimmer\\
    \midrule
    \multirow{3}{*}{CPG} & Topic 14: Visual appeals & power\\
    & Topic 16: Humor & peak\\
    &Topic 12: Visual aesthetics & ddp\\
    \midrule
    \multirow{3}{*}{Automobile} & Topic 12: Visual appeals & maximum pitch\\
    & Topic 15: Realism & tempo\\
    & Topic 2: Storytelling & power\\
    \midrule
    \multirow{3}{*}{Entertainment} & Topic 14: Entertainment & dB \\
    & Topic 8: Endorsement / celebrity & average pitch\\
    & Topic 16: Humor & maximum pitch\\
    \bottomrule
\end{tabular}
\end{table}

\begin{figure}[ht]
    \centering
    \includegraphics[width=1\linewidth]{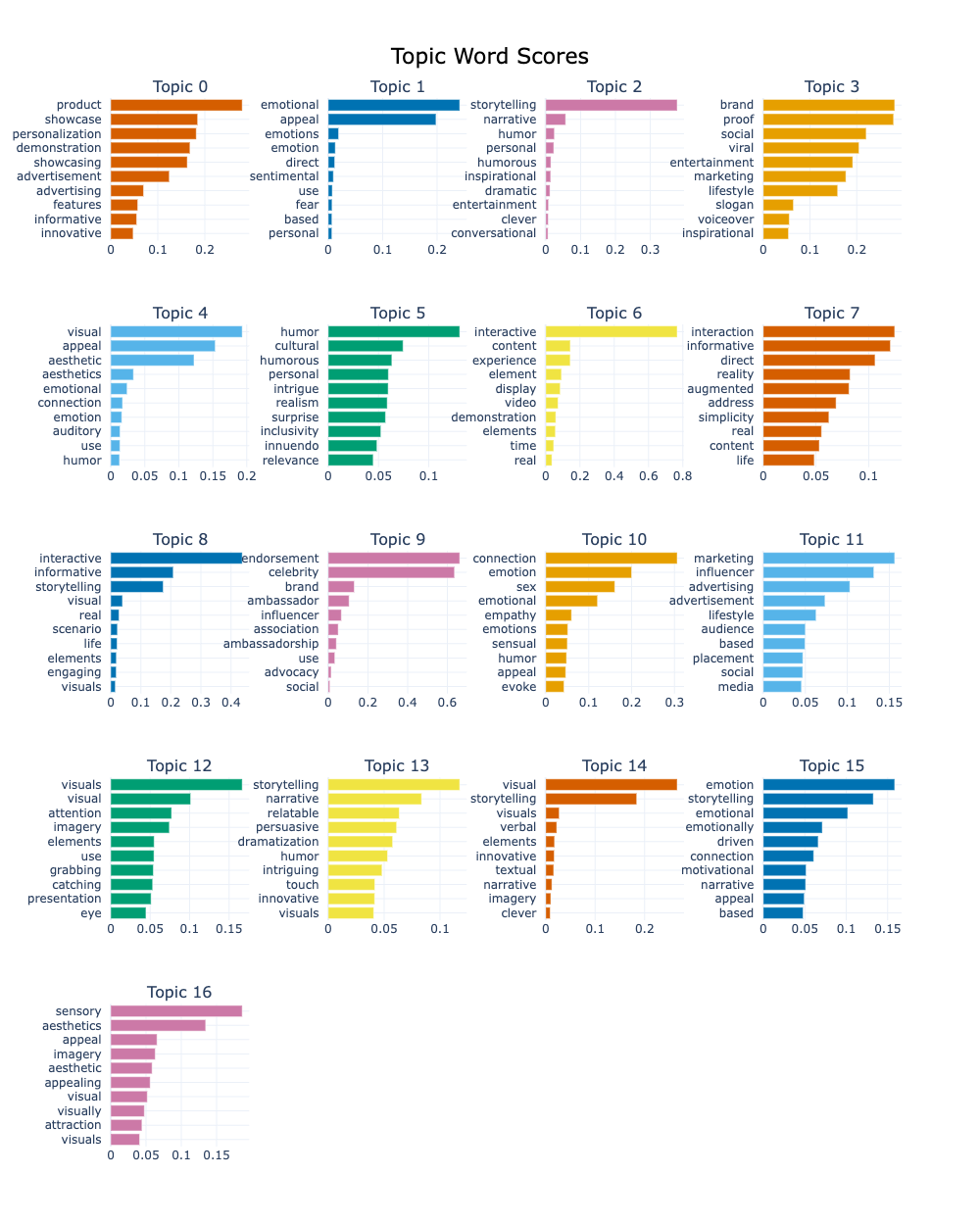}
    \caption{Topics of hooking period design methods for videos in Ecommerce. Only top 10 words are shown for each topic}
    \label{fig:ecomm_topics}
\end{figure}

\textbf{Feature Importance} Table \ref{tab:top_features} presents the top topics and acoustic features of video hooking periods that correlate with CPI. Specifically, ``Interactive content" emerges as the most effective design methodology for engaging audiences in Ecommerce video ads, followed by ``Interaction" and ``Connection." However, effectiveness varies notably across different verticals. For instance, ``Demo/Product" is the leading methodology in Healthcare, whereas ``Visual appeals" excels in the Consumer Packaged Goods sector. The topics were derived using BERTopic, based on the design methodology and accompanying reasoning illustrations provided by the vision design methodology extractor. Overall, we identified 17 key methods/topics frequently employed by advertisers to engage their audience\footnote{The number of topics (e.g., 17) is chosen based on the relatively optimal perplexity score among a few choices.} (see the Ecommerce topics in Fig. \ref{fig:ecomm_topics} for details).

\begin{figure*}[ht]
    \centering
    \subfigure{\includegraphics[width=0.47\linewidth]{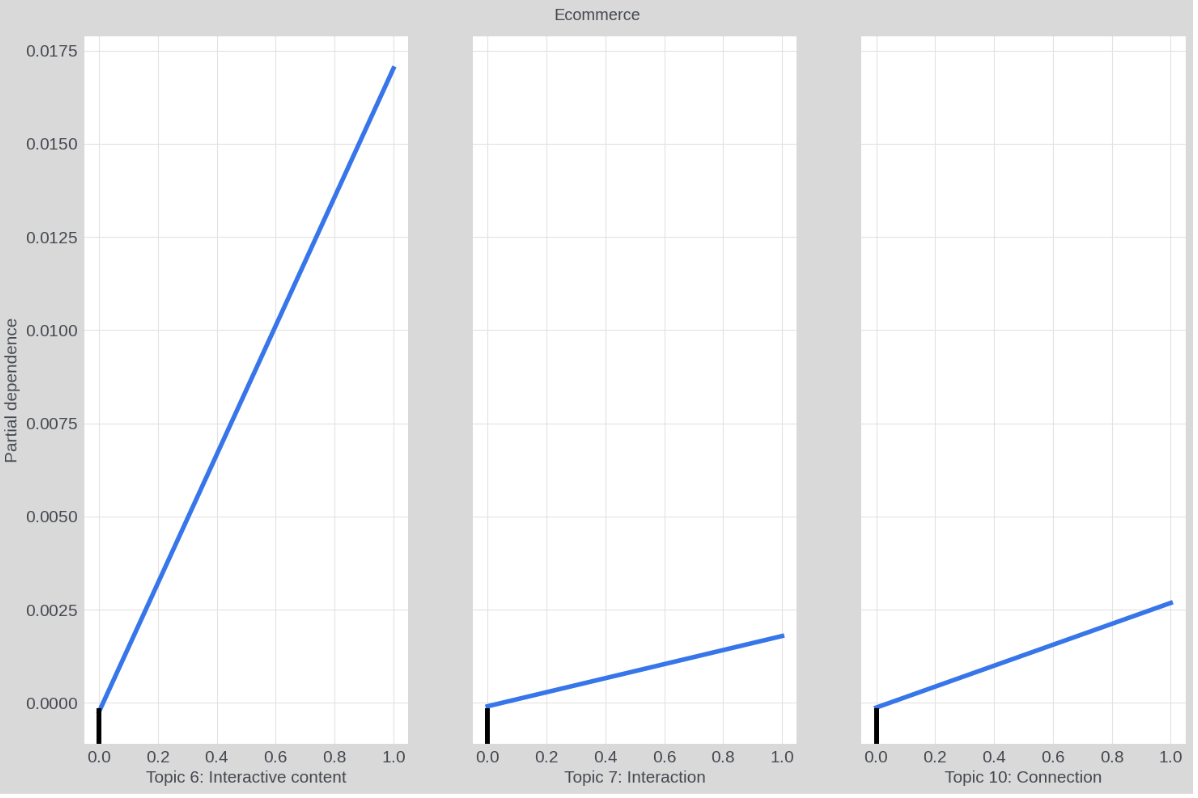}}
    \subfigure{\includegraphics[width=0.47\linewidth]{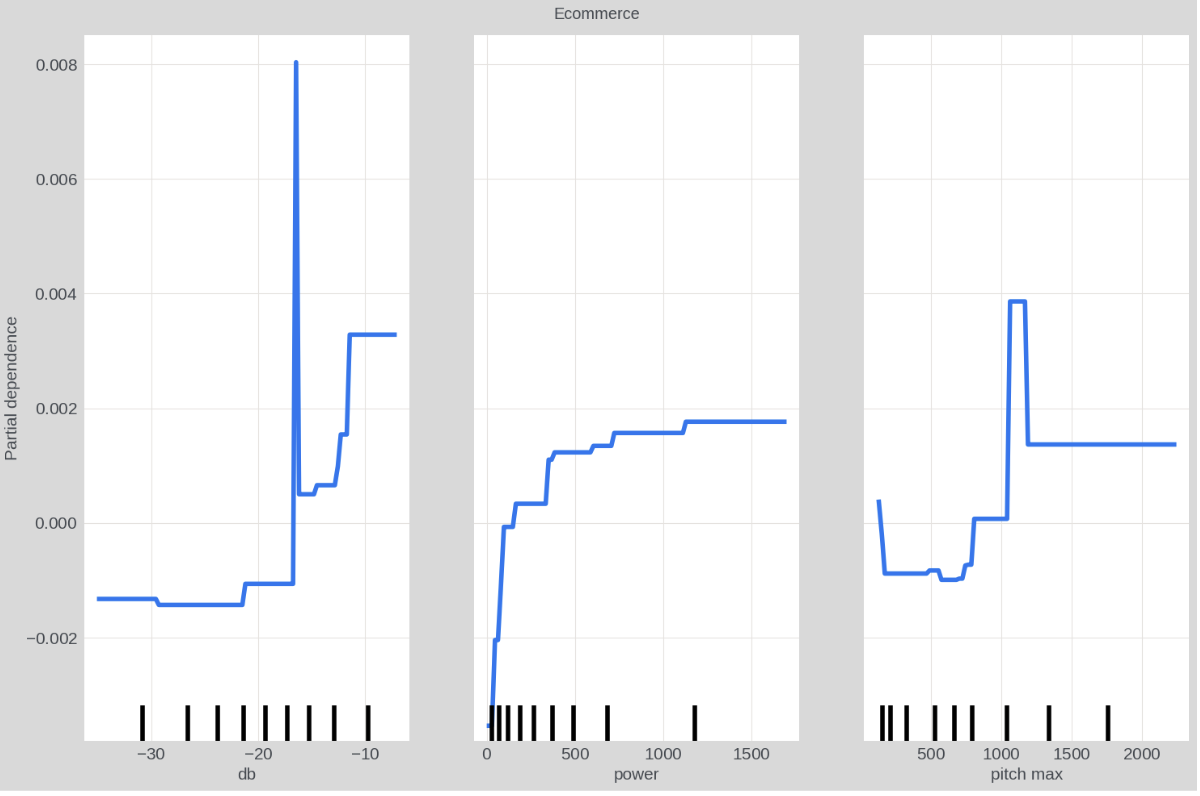}}
    \caption{PDP for top 3 visual topics (left) and top 3 acoustic features (right) for Ecommerce}
    \label{fig:ecomm_pdp}
\end{figure*}

\textbf{Partial Dependence} To further examine the relationship between a specific feature and the predicted outcome of our model while controlling for the effects of other features, we present partial dependence plots (PDPs) \cite{pdp2001}. An upward trend in the PDP line indicates a positive relationship, meaning the prediction increases as the feature value increases. Conversely, a downward trend indicates a negative relationship. Figure \ref{fig:ecomm_pdp} displays the PDPs for the top three visual and acoustic features, respectively (similar plots for other verticals are provided in the Appendix).

The visual feature analysis suggests that incorporating more interactive content during the hooking period of Ecommerce videos can effectively increase CPI, exerting a stronger influence compared to interaction and connection design elements. In contrast, the acoustic features from topic 3 demonstrate non-linear relationships. Specifically, an optimal range for decibel levels (dB) and maximum pitch can lead to higher CPI, whereas the power feature shows a threshold effect.

\section{Conclusion}
In conclusion, this paper presents a framework for systematically analyzing the hooking period of video advertisements through advanced multimodal large language models. Our framework effectively leverages transformer-based architectures to integrate visual, acoustic, and textual features, enabling a comprehensive understanding of the nuanced interplay among these multimodal elements. The empirical evaluation conducted on large-scale, real-world data demonstrates that our approach can significantly enhance the prediction of advertisement performance metrics, notably the Conversion Per Investment (CPI).

The findings from our analysis provide actionable insights that can guide advertisers in developing more effective and targeted video ad strategies tailored to specific industry verticals. By identifying important features that influence viewer engagement during the initial moments of ads, our method empowers advertisers to optimize content strategically and maximize viewer attention and subsequent performance outcomes. Future research could further expand this framework by incorporating additional data modalities such as viewer emotional responses, gaze tracking, or user interaction metrics, and exploring more sophisticated multimodal fusion techniques to enhance predictive accuracy and generalizability.

Despite the promising results, our study is not without limitations that present opportunities for future research. First, the analysis is confined to the initial three seconds of a video ad, which may not capture the full dynamics of viewer engagement across the entire video. Second, the reliance on pretrained multimodal large language models introduces potential biases and sensitivity to prompt design. Furthermore, the dataset, though extensive, is platform-specific and thus may not generalize to other environments. Addressing these limitations in future work would further strengthen the robustness and applicability of the proposed framework.

Although our system is designed to extract actionable insights from video-based ads using Multimodal Large Language Models (MLLMs), deployment to live users has been blocked due to real-world challenges. Specifically, regulatory constraints around user privacy and ad targeting have prevented us from launching the system at scale. We believe that demonstrating the benefit of this approach can motivate both scholars and practitioners to further explore LLM applications for multimodal data. We provide detailed documentation of our deployment attempts and the key barriers encountered.

\newpage
\bibliographystyle{ACM-Reference-Format}
\bibliography{ref}

\newpage
\section*{Appendix: Showcase} \label{sec:showcase}
\begin{figure*}[h!]
     \centering
       \captionsetup{justification=raggedright,singlelinecheck=false}
     \subfigure{\includegraphics[width=0.259\textwidth]{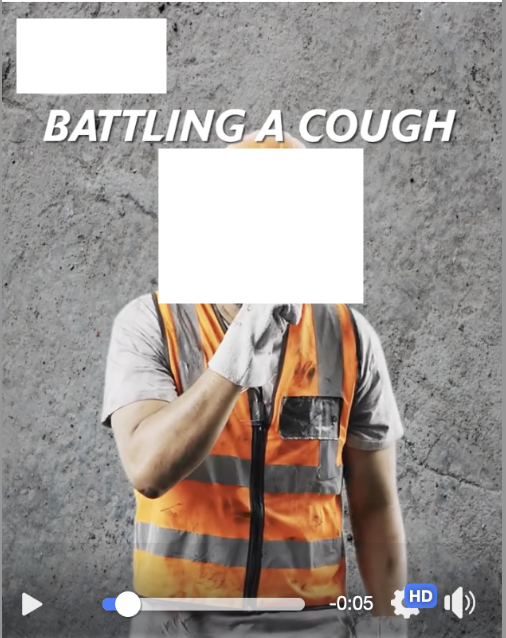}} 
     \subfigure{\includegraphics[width=0.32\textwidth]{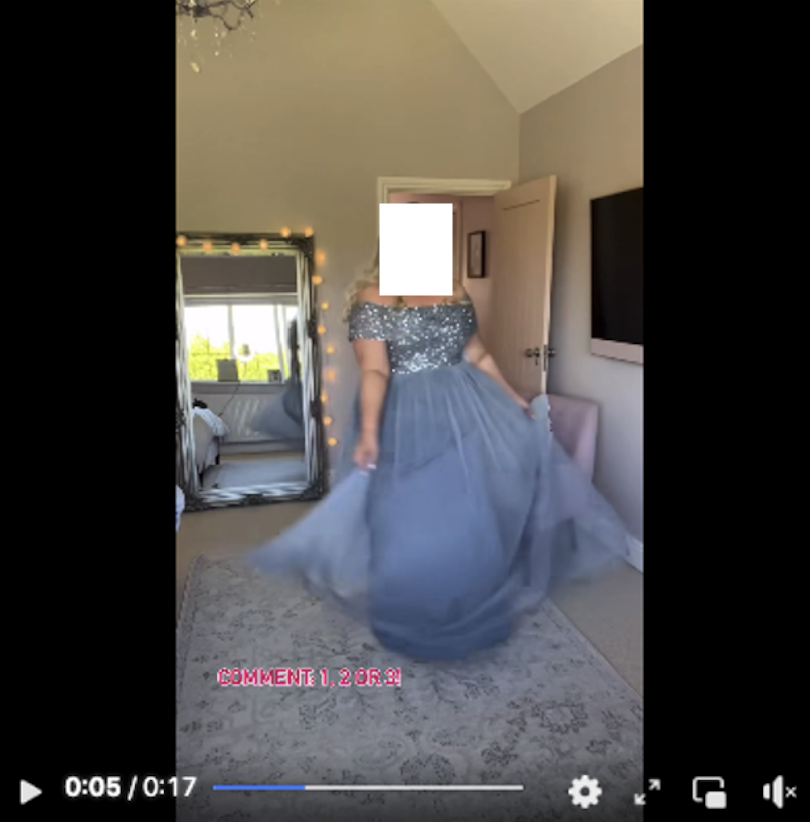}} 
     \subfigure{\includegraphics[width=0.26\textwidth]{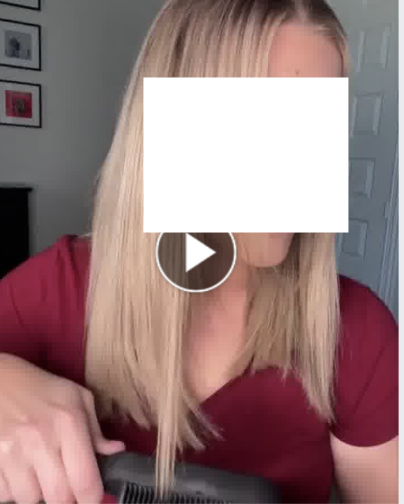}}\\
     \subfigure{\includegraphics[width=0.226\textwidth]{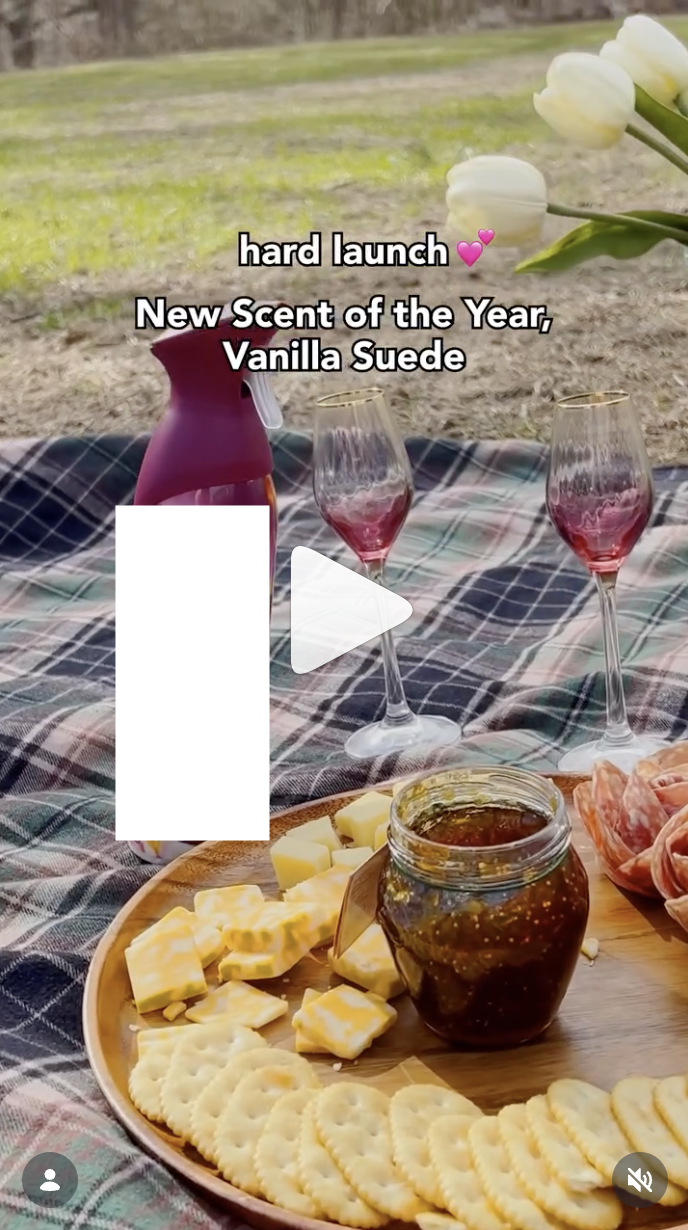}} 
     \subfigure{\includegraphics[width=0.32\textwidth]{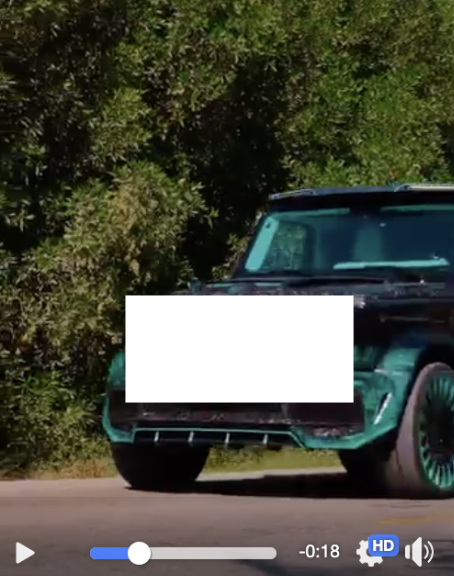}}
     \caption{Top row: Example 1 (Healthcare, @benylinsa): https://www.instagram.com/benylinsa/reel/C8667kyKkNS, \newline Example 2 (Entertainment, @mayadeluxeuk):https://www.instagram.com/mayadeluxeuk/reel/C7WiUL2tMAv, and  \newline Example 3 (Ecommerce, @momglamlife): https://www.facebook.com/reel/644847075355183; Bottom row: \newline Example 4 (Consumerm @proctergamble):https://www.instagram.com/reel/DLCbpbjI55Y and \newline Example 5 (Automobile, @douradoluxurycars):https://www.instagram.com/reel/DEz6LLdtDGw}
     \label{fig:example}
 \end{figure*}

 We showcase some of the top-performing video ads from our extensive dataset (see Fig. \ref{fig:example}).
 These examples are public ads that can be seen via the link provided, and we are showing here mockups to respect the creative rights and these serve to provide face validity for the effectiveness of our method, particularly emphasizing the robustness of our feature extraction process facilitated by MLLM. By analyzing these standout ads, we can better understand the elements that contribute to their success and how our approach successfully identifies and leverages these key features. Example 1 highlights a human action centered around demonstrating a coughing symptom with the produce included in the initial hooking period of the advertisement. This relatable activity effectively captures the audience's attention right from the start, establishing an emotional connection that engages viewers and encourages them to continue watching. Example 2 focuses on the use of celebrity endorsement or influencer collaboration, where the featured individual is seen posing in a manner that resonates with the target audience. The presence of a well-known personality not only enhances the ad's credibility but also leverages the influencer's existing fan base to broaden the advertisement's reach and impact. This strategic positioning helps in building trust and persuading viewers through association with a familiar and respected figure. Example 3 revolves around a promotional strategy that incorporates a direct response. This approach is designed to prompt immediate viewer interaction, e.g., making a purchase, or visiting the retailer's website. By clearly communicating the desired action, the ad effectively drives conversions and achieves specific marketing objectives, demonstrating the practical application of persuasive techniques within the promotional content. Example 4 is the ad from P\&G, showing some humor (including a message in the first 3 seconds: ``hard launch $\heartsuit$ New Scent of the Year, vanilla Suede") with a visual appealing background (e.g., green plants and wall art decoration). Example 5 features a video introducing a conceptual luxury car through a narrative-driven presentation, telling a story that highlights the car's design philosophy and innovative features.

 These examples collectively illustrate the diverse strategies employed in successful video advertisements. They also demonstrate how our method accurately identifies and extracts critical features that contribute to high performance. By leveraging multimodal LLMs for feature extraction, our approach not only validates the effectiveness of these advertising techniques but also provides a scalable framework for analyzing and optimizing future video ad campaigns.

\end{document}